\documentclass{bcp_arch}
\usepackage{graphics}
\usepackage{graphicx}
\usepackage{amsmath}

\def\LaTeX{L\kern -.36em\raise .3ex\hbox{\sc a}\kern -.15em T\kern -.1667em%
\lower .7ex\hbox{E}\kern -.125em X}

\font\tt=cmtt10
\begin{document}

\mathclass{Primary 92D25; Secondary 82-06.}
\thanks{M.M. gratefully acknowledges the support of the German Alexander von Humboldt Foundation through the Fellowship No. IV-SCZ/1119205 STP.}
\abbrevauthors{T. Reichenbach et al.}
\abbrevtitle{Stochasticity in cyclic coevolutionary dynamics }
  
\title{Stochastic effects on biodiversity \\ in cyclic coevolutionary dynamics}

\author{Tobias Reichenbach, Mauro Mobilia, and Erwin Frey}
\address{Arnold Sommerfeld Center for Theoretical Physics
(ASC) and
  Center for NanoScience (CeNS),\\ Department of Physics,
  Ludwig-Maximilians-Universit\"at M\"unchen,\\ Theresienstrasse 37,
  D-80333 M\"unchen, Germany\\
  {\tt E-mail:tobias.reichenbach@physik.lmu.de,mauro.mobilia@physik.lmu.de,frey@lmu.de}}

\maketitlebcp

\abstract{Finite-size fluctuations arising in the dynamics of competing populations may have dramatic influence on their fate. As an example, in this article, we investigate a model of three species which  dominate each other in a cyclic manner. Although the deterministic approach predicts (neutrally) stable coexistence of all species, 
for any finite population size, the intrinsic stochasticity unavoidably causes the eventual extinction of two of them.}    

\section*{1. Introduction.} Cyclic dominance of species has e.g. been observed in lizard populations [Sin]
as well as in bacterial communities [Kerr]. Hereby,  three different morphisms of lizards resp. three
strands of bacteria compete in a non-transitive way: each morphism/strand outperforms another but is
in turn dominated by the third.  The resulting cyclic coevolutionary dynamics may feature
biodiversity [Kerr], which constitutes a central topic in modern ecology [May, Neal] and environmental
policy [CNRS]. Much interest and effort is therefore devoted to the theoretical understanding of
such systems.
In the context of evolutionary game theory [Hof], the paradigmatic model for cyclic coevolution is
the well-known rock-paper-scissors game, where each strategy beats another but is itself defeated by
a third. The replicator dynamics deterministically describes the system's time-evolution, yielding
oscillatory behavior.  Similar results were already obtained in the pioneering work by Lotka [Lot]
and Volterra [Vol], who described fish catches in the Adriatic through nonlinear differential equations,
exhibiting oscillations as well. Despite their success and popularity, as a common shortcoming, these 
approaches ignore any form of internal noise arising from 
size fluctuations, which are  unavoidable for finite populations,
as well as from the stochastic spatial distribution of the reactants. In this article, we will
specifically focus on the effects of finite-size fluctuations on the coevolution of a three-species
model. Actually, the deterministic approaches (tacitly) assume that any finite-size effects should disappear
in the limit of large populations, where they are expected to become valid.
Only recently, the influence of the finite character of populations, as well as the relation between 
stochastic models and their deterministic counterparts have been investigated (see e.g. [Tra1, Tra2]). \\
In this article, we consider a paradigmatic microscopic model for cyclic coevolutionary dynamics and study the effects of internal stochasticity as compared to the deterministic predictions. We show that finite-size fluctuations dramatically alter the system's behavior: While the rate equations predict the existence of (neutrally) stable cycles, for any size, stochasticity leads the system to the extinction of two species. Technical details of our analysis can be found in [Rei], here we focus on a more general and intuitive discussion.

\section*{2. The model.} We consider three populations, labeled as $A$, $B$, and $C$, which mutually inhibit each other: $A$ invades $B$, $B$ outperforms $C$, and $C$ dominates over $A$, closing the cycle. For the stochastic dynamics, we consider a finite number $N$ of overall individuals in an ``urn'' [Fel], i.e. in a well-mixed environment. Two individuals may react with each other at certain rates $k_A,~k_B$ and $k_C$ according to the following reaction scheme:
\begin{equation}
A+B\stackrel{k_C}{\longrightarrow} A+A~,\quad
B+C\stackrel{k_A}{\longrightarrow} B+B ~,\quad
C+A\stackrel{k_B}{\longrightarrow} C+C~ .
\label{react} 
\end{equation}
Such ``urn models'' are closely related to the Moran process [Mor], which describes stochastic evolution of finite populations with constant fitness.\\
Note that since the reactions (\ref{react}) conserve  the total number $N$ of individuals, one is left only with two degrees of freedom exist.

\section*{3. Deterministic analysis.} 
\begin{figure}
\begin{center}
\includegraphics[scale=0.85]{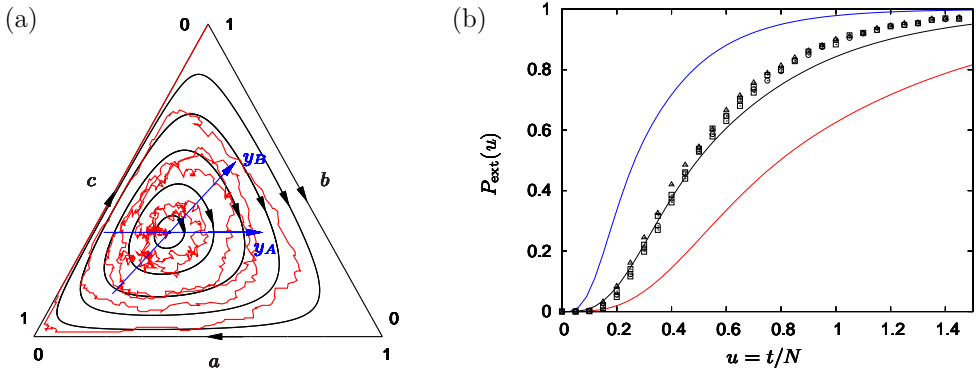}
\caption{(color online) Reproduced from [Rei]. The three-species simplex for reaction rates $k_A=1,~k_B=1.5,~k_C=2$ is drawn in (a). The rate equations predict cycles, which are shown in black. Their linearization around the reactive fixed point is solved in proper coordinates $y_A,y_B$ (blue or dark gray). The red (or light gray) erratic flow is a single trajectory in a finite system ($N=200$), obtained from stochastic simulations. It spirals out from the reactive fixed point, eventually reaching an absorbing state.
The extinction probability $P_{\rm ext}$ is reported in (b)
as function of the rescaled time $u=t/N$. Numerical data are shown for different system sizes $N=100$ (triangles), $200$ (boxes), $500$ (circles). The red (light gray), blue (dark gray), and black curves correspond to analytical results for different values  $R=1/\sqrt{3}$ (top), $~1/3$ (bottom), and their average value (middle). 
\label{simplex_ext}}
\end{center}
\end{figure}

Let us start with the deterministic approach, in which
the mean-values of the densities of species $A,~B$, $C$, denoted  respectively $a,~b$, and $c$,  evolve according to the so-called rate equations, which are easily obtained and read:
\begin{equation}
\dot{a}=a(k_Cb-k_Bc)~, \quad
\dot{b}=b(k_Ac-k_Ca)~,\quad
\dot{c}=c(k_Ba-k_Ab)~.
\label{RE}
\end{equation}
Due to the conservation of the total density (for commodity, we set $a+b+c=1$ ) only two of the three equations are independent. In addition, the quantity $K=a(t)^{k_A}b(t)^{k_B}c^{k_C}$ is conserved by Eqs.~(\ref{RE}) [Rei] and, as for mechanical systems with conserved energy, is responsible for cyclic orbits in the phase space.
In Fig.~\ref{simplex_ext} (a), the solutions of Eqs.~(\ref{RE}) are shown in the two-dimensional phase portrait (simplex) for different initial conditions. The flows are cyclic trajectories around the neutrally stable 
reactive fixed point $(a^*,b^*,c^*)=(k_A,k_B,k_C)/(k_A + k_B + k_C)$, which corresponds to a coexistent state
with nonvanishing densities of all species. On the contrary, the corners of the simplex 
are associated with absorbing states, where only one species survives. In addition, the boundary of the simplex, is absorbing: once it has been reached, one of the species gets extinct and, among  the two remaining populations, one dominates the other and the system is driven to one of the corners. As a consequence of 
the conservation of $K$, 
\begin{equation}
\mathcal{R}=\mathcal{N}\sqrt{a^{*k_A}b^{*k_B}c^{*k_C}-K}\quad,
\label{const2}
\end{equation}
is also left invariant by Eqs.~(\ref{RE}). Here, $\mathcal{N}$ is a normalization factor. 
As any perturbations to one closed orbits of  Fig.~\ref{simplex_ext} (a)
give rise to a different regular cycle, they latter are said to be \emph{neutrally stable}.

The quantity $\mathcal{R}$ monotonically increases with the distance from the steady state where it vanishes. Hence, it provides an efficient measure of the separation length between the reactive fixed point and any deterministic cycles. Furthermore, in the vicinity of the stationary state, the linearization of Eqs.~(\ref{RE})
gives an accurate description of those orbits, which become circles in well suited 
coordinates ($y_A,~y_B$ in Fig. \ref{simplex_ext} (a)). In the linear regime, $\mathcal{R}$ 
is  the radius of these circles whose characteristic frequency is $\omega_0=\sqrt{\frac{k_Ak_Bk_C}{k_A+k_B+k_C}}$ [Rei].

\section*{4. Stochastic behavior.}

The stochastic model does not conserve the quantity (\ref{const2}) and, as a consequence, 
the cyclic structure of the flows is lost. In fact, when one takes internal noise into account,
the trajectory in the phase space can jump between the deterministic cycles. As an illustration of this fact,
in Fig. \ref{simplex_ext} we have reported one of such a stochastic flow. Here, starting from the reactive fixed point, the trajectory soon departs from the latter,
spirals around erratically and deviate from it with time.
At some point, such a flow hits the absorbing boundary and ends 
up in one of the absorbing states. This is the generic
 scenario followed by the system. Indeed, due to the neutrally stable character of
the deterministic cycles, the stochastic trajectory can perform a `random walk' in the phase portrait 
between  them. Eventually, the 
absorbing boundary is reached and coexistence of all three species is lost. Hence, stochasticity arising from 
finite-size effects causes extinction of two species.
In the following, we outline an analytical approach allowing for a quantitative description
of the above behavior, with a proper treatment of the finite-size fluctuations
(see [Rei] for a detailed discussion).
 For the sake of simplicity, we assume equal reaction rates ($k_A=k_B=k_C=1$) in the basic reactions (\ref{react}) that define the stochastic process. As usual, the dynamics of the latter is
 enconded in the related master equation. From the latter, it is fruitful to proceed with
a Kramers-Moyal expansion (see e.g. [Gar]) which leads, up to the second term, to a 
Fokker-Planck equation (FPE). To make further progress, we consider a small noise approximation
and linearize the FPE around the reactive fixed point [Rei].  We also exploit the above-mentioned symmetry 
of the deterministic cycles in the $y$-variables, shown 
in Fig. \ref{simplex_ext}, and introduce the polar coordinates $r,~\phi$. In the latter, the FPE
is recast in the neat form
\begin{equation}
\partial_t P(r,\phi,t)=-\omega_0\partial_\phi P(r,\phi,t)
+\frac{1}{12N}\Big[\frac{1}{r^2}\partial_\phi^2+\frac{1}{r}\partial_r+\partial_r^2\Big]P(r,\phi,t)~.
\label{pol-f-p}
\end{equation} 
The interpretation of this equation is simple and enlightening.
The first term on the right-hand-side of (\ref{pol-f-p}) 
accounts for the deterministic oscillatory dynamics with frequency 
$\omega_0$. The second term, which is a Laplacian operator in polar coordinates and is proportional to $1/N$, 
encodes the finite-size fluctuations and ensures the (isotropic) diffusive
character of the dynamics. It tells that fluctuations decrease with the system size, $N$.\\
Ignoring the absorbing character of the boundary for a moment, the FPE (\ref{pol-f-p}) 
allows us to compute the mean-square displacement $\langle r^2(t) \rangle$ from the reactive fixed point, 
when initially starting from there. In this situation, the initial probability distribution is a delta peak at 
the reactive steady state, which broadens in time, remaining a spherically symmetric gaussian 
distribution. Eventually, we obtain $\langle r^2(t) \rangle=t/(3N)$. Thus, larger systems necessitate a longer waiting time to reach a given deviation from the reactive fixed point. 

\section*{5. Extinction probability.} Taking the absorbing nature of the boundary into account, we now investigate the extinction probability $P_\text{ext}(t)$, which is the probability at time $t$ that the system, which was initially in its reactive steady state, has reached the absorbing boundary and two species
become extinct. Note that an initially spherically symmetric probability distribution, as a delta peak in our case, is left spherically symmetric by the FPE (\ref{pol-f-p}). The dependence on $\phi$ in the latter equation therefore drops out, and only the second term remains, describing isotropic diffusion. Finding $P_\text{ext}(t)$ thus translates into a first-passage problem to a sphere. Note that, due to our linearization at the reactive fixed point, the triangular-shaped boundary is mapped onto a sphere, which is of course inaccurate. However, we are able to account for nonlinearities in a pragmatic manner, as shown below. The solution to the first-passage problem to a sphere is well-known and may e.g. be found in [Red]. We therefore directly obtain the Laplace transform of $P_\text{ext}$, which reads
\begin{equation}
\text{LT}\{P_\text{ext}(u)\}=\frac{1}{sI_0(R\sqrt{s/12})}
\label{ext_prob_u}
\end{equation}
where $u=t/N$ is the rescaled time. As already found before, increasing the system size $N$ only has the 
effect of protracting the time at which extinction occurs. In Fig. \ref{simplex_ext} (b) we compare results from 
stochastic simulations (based on the Gillespie algorithm) to our analytical approximations. For the 
simulations, different total populations sizes $N$ were considered and, 
when rescaling time according to $u=t/N$, the data are seen to collapse on a universal curve, validating the scaling result. As for the analytical curves, nonlinear effects were dealt with 
by considering different 
distances $R$ to the absorbing boundary. Small/large values of $R$ over/under-estimate the numerical results. However, for an intermediate value of $R$, we obtain an excellent 
agreement.

\section*{6. Conclusions.} We have presented a simple stochastic model for cyclic coevolutionary dynamics. Biodiversity in the form of coexistence of all three species emerges within the deterministic approach, which leads to neutrally stable oscillations. As the fluctuation drive the system to extinction of two species, stochasticity invalidates the rate equations predictions. We have quantified this behavior by deriving an 
appropriate Fokker-Planck equation, using linearization around the reactive steady state and by exploiting the polar symmetry of the system.

\references{CNRS}
{ 
 
\item{[CNRS]} Le journal du CNRS, {\bf N. 196} (2006)

\item{[Fel]} W. Feller, {\it An Introduction to Probability Theory and its Application}, 3rd ed., Vol. 1, Wiley, New York, 1968. 

\item{[Gar]} C. W. Gardiner, {\it Handbook of Stochastic Methods}, 1st ed., Springer, Berlin, 1983.

\item{[Hof]} J. Hofbauer and K. Sigmund, {\it Evolutionary Games and Population Dynamics}, Cambridge University Press, Cambridge, 1998.

\item{[Kerr]} B. Kerr, M.~A.Riley, M.~W. Feldman, and B.~J.~M. Bohannan, {\it Local dispersal promotes biodiversity in a real-life game of rock-paper-scissors},   Nature
  {\bf 418} (2002), 171.

\item{[Lot]}    A. J. Lotka, J. Amer. Chem. Soc. {\bf 42} (1920), 1595.
  
\item{[May]} R. M. May, {\it Stability and Complexity in Model Ecosystems}, Cambridge University Press, Cambridge, 1974.

\item{[Mor]} P. A. P. Moran, {\it Random processes in genetics}, Proc. Camb. Phil. Soc. {\bf 54} (1958), 60.

\item{[Neal]} D. Neal, {\it Introduction to Population Biology}, Cambridge University Press, Cambridge, 2004.
  
\item{[Red]} S. Redner, {\it A guide to first-passage processes}, Cambridge University Press, Cambridge, 2001 .

\item{[Rei]} T. Reichenbach, M. Mobilia, and E. Frey, {\it Coexistence versus extinction in the stochastic cyclic Lotka-Volterra model}, accepted in Physical Review E, {\tt q-bio.PE/0605042}.

\item{[Sin]} B. Sinervo B. C. M. and Lively, {\it The rock-scissors-paper game and the evolution of alternative male strategies}, Nature {\bf 380} (1996), 240.
  
\item{[Tra1]} A. Traulsen, J. C. Claussen, and C. Hauert, {\it Coevolutionary Dynamics: From Finite to Infinite Populations}, Phys. Rev. Lett. {\bf 95} (2005), 238701.

\item{[Tra2]}  A. Traulsen, J. C. Claussen, and C. Hauert, {\it Coevolutionary dynamics in large, but finite populations}, Phys. Rev. E {\bf 74} (2006), 011901.

\item{[Vol]} V. Volterra, Mem. Accad. Lincei {\bf 2} (1926), 31.

}

\end{document}